# The feasibility of launching physical layer attacks in visible light communication networks

*Grzegorz Blinowski*
*Email: g.blinowski@ii.pw.edu.pl*
*Institute of Computer Science,*
*Warsaw University of Technology, Nowowiejska 15/19, 00-665 Warszawa, Poland*

**Abstract**

One of the areas in which wireless networks based on visible light communication (VLC) are considered superior to traditional radio-based communication is security. The common slogan summarizing VLC security features is: WYSIWYS - "What You See Is What You Send". However, especially in the case of infrastructure downlink communication, security with respect to data snooping, jamming and modification must be carefully provided for. This paper examines the physical layer aspects of VLC networks with respect to possible disruptions caused by rogue transmitters. We present the theoretical system model that we use in simulations to evaluate various rogue transmission scenarios in a typical office environment. We use estimated Bit Error Rate (BER) as a measure of the effectiveness of jamming and rogue data transmission. We find that it is quite easy to disrupt, and in some cases to even hijack legitimate transmission.

## 1. Introduction

Visible light communication (VLC) is a wireless optical communication technology through which baseband signals are modulated on the light emitted by an LED: [1] – [3]. The decreasing cost and hence rapid adaptation of LED-based light make VLC a promising communication technique and a significant alternative to radio-based wireless communication. As user demand for data transmission throughput and availability continues to increase, "traditional" radio-based communication systems, such as Wi-Fi, Bluetooth, ZigBee, etc. fail to deliver because of their bounded channel capacity and transmission rates due to the limited radio spectrum available. VLC data transmission networks provide an attractive alternative to traditional wireless techniques.

VLC systems have been proposed and implemented both for indoor and outdoor applications – see [2] and [4]. Indoor applications include a range of communication facilities provided today by radio-based WLAN and Personal Area Networks (PAN), and range from: office communication – [5], multimedia conferencing – [6], peer-to-peer data exchange, data broadcasting – especially multimedia such as home-audio and video streams, see: [7] – [10], to positioning: [11], [12]. VLC systems also provide a safe alternative to electromagnetic interference from radio frequency communications in hazardous environments, such as mines and petrochemical plants, and in applications where traditional WLAN communication may interfere with specialized equipment, for example in hospitals and aeronautics [13].

VLC is also starting to be considered as a way of augmenting or even replacing RF networks; for example, a wide range of techniques aimed at VLC based multimedia networks was developed under "hOME Gigabit Access" project (OMEGA) [14] sponsored by the European Union. The usage of smartphone cameras and light sensors brings VLC to the field of mobile computing and sensing. VLC has the potential to evolve into a general WLAN standard – in [15] with the OpenVLC platform, the authors have demonstrated that it is relatively easy with current Software Defined Radio (SDR) toolkits to implement the TCP/IP suite on the VLC medium.

One of the areas in which VLC techniques are considered superior to traditional radio-based communication is security. The directivity and high obstacle impermeability of optical signals are considered to provide a secure way to transmit data within an indoor environment, making the data difficult to intercept from outside. The common slogan summarizing VLC security features is: WYSIWYS – "What You See Is What You Send" [16]. As the recent history of IT technological progress has taught us, a common mistake in the development of novel communication techniques has been to ignore or marginalize security issues. Such was the case with the IPv4 internet protocol suite, fiber-optics based networks, and more recently, with early adopted WLAN technologies. Currently, the VLC industry seems to be on the same path again: the indubitable "pro-security" physical characteristics of visual light communication have steered the developers' focus away from the security track.

The shared nature of the medium allows wireless networks to be easily monitored and broadcast on. Attackers may not only easily monitor communication but also launch jamming (denial of service) attacks. Attacks on the physical level that disregard MAC-level protocols can effectively block the network and are not remedied by traditional security mechanisms. A risk assessment of VLC communication with respect to the communicating parties of three basic classes: mobile (smartphones, tablets, wearables, etc.), fixed (PCs, peripherals, and appliances) and infrastructure (fixed in-room transmitters) was conducted in [17]. By analyzing basic physical characteristics of the VLC communication channel, it was shown that particularly in the case of infrastructure downlink communication, security with respect to data snooping, jamming and modification must be carefully provided for. In order to ensure the dependability of VLC networks, we must better understand the physical layer mechanisms of executing a hostile-transmitter type attack. This paper examines the physical layer aspects of VLC networks with respect to possible disruptions caused by rogue transmitters.

The structure of this work is as follows: in section 2, we will describe the current state of VLC security research; in section 3, we will present the theoretical system model which we will use in subsequent simulations to evaluate various rogue transmission scenarios; in section 4, we will describe the physical properties of evaluation scenarios; in section 5, we will show and discuss simulation results; and in section 6, we will summarize the paper and outline areas of future research.

## 2. Current state of VLC security research

### 2.1. Confidentiality and data snooping

A common assumption in VLC, as stated for example in the IEEE 802.15.7 standard [18], is that: "Because of directionality and visibility, if an unauthorized receiver is in the path of the communication signal, it can be recognized." However, this is not always true; let us consider a case of communication with an "infrastructure" transmitter. Both in the case of the NLOS channel and LOS, an unauthorized receiver may be easily introduced into the environment without being recognized.

Snooping on VLC transmission is, of course, limited by high obstacle impermeability and is more difficult than Wi-Fi snooping, but there is no obvious reason why it should not be possible. In [19], it was shown experimentally that eavesdropping on VLC transmission is indeed possible. The equipment used, based on a standard low-cost SDR design, was able to achieve acceptable BER rates in a range of different scenarios. The authors evaluated different room configurations and were able to decode high-order modulated 64-QAM VLC signals outside of the room, via door gaps, key holes and windows protected by special "privacy" coatings.

Up till now, the confidentiality of VLC communication has mainly been tackled on the information-theory level, referring to the discrete memoryless wiretap channel and the metric of the channel's secrecy capacity, as originally introduced by Wyner [20]. In [21], the authors proposed to use the MIMO technique and beam-forming (similar to BF/MIMO implemented in Wi-Fi networks) to establish a secure channel between the transmitter and the receiver located in a particular physical location. The BER (Bit Error Rate) was minimized at the receiver's location, while it remained unacceptably high in the rest of the area. In this way, a potential eavesdropper physically located some distance from the legitimate receiver was unable to properly decode the data. This was attained without significant influence on the lighting characteristics and was therefore unobservable to the users. A similar approach was proposed in [22] using the MISO (Multiple Input Single Output) technique, together with null-steering and artificial noise – an achievable secrecy rate was calculated numerically. In a related work [23], this approach was also in part verified in a real environment.

### 3. The system model

The system components are an LED transmitter, consisting of multiple light sources, and a photodiode receiver. The

### 2.2. Jamming and data modification

Let us consider the possible schemes for introducing a signal jamming or data-modifying device into the VLC infrastructure channel. The attacker may choose to use both a directed - Line of Sight (LOS) or non-directed Non-Line of Sight (NLOS) transmitter-receiver arrangement; but, due to power considerations, a LOS model will be preferred. In general, the attacker's aim is to achieve a higher illumination at the receiver than that provided by the legitimate transmitter.

The major practical concern from the attacker's point of view is to ensure that the illumination provided by the rogue transmitter remains undetected by users. Hence, the attacker may introduce his own (preferably) highly directed transmitter or "hijack" a part of the legitimate infrastructure. VLC infrastructure networks consist of numerous independent transmitters to provide adequate coverage and capacity. Multi-transmitter "femtocell" VLC networks are also studied as an extension of traditional Wi-Fi and cellular networks [24]. In such environments, the installation of a rogue transmitter may easily pass undetected. A second possibility is hijacking a part of the legitimate VLC infrastructure via a wired or wireless channel. In a large installation, such malicious intervention may easily pass undetected.

Data modification in VLC networks may be attained by reactive jamming techniques. As was demonstrated in [25], real time reactive jamming is easily in reach of attackers with the use of SDR technology. In the above mentioned work, ZigBee (IEEE 802.15.4) protocol devices were used – it is worth noting the MAC-level similarities of ZigBee and the VLC 802.15.7 standard.

received signal depends on the physical characteristics of the transmitting LED, the receiver, and channel characteristics which, in turn, are determined by a room's physical properties (dimensions, wall reflectivity). We

use ray optics to calculate signal and noise levels and to derive adequate metrics. We assume the MISO (Multiple Input Single Output) model with multiple luminaires, each consisting of multiple transmitting LEDs (both legitimate and rogue) and one photodiode detector. A single transmitting LED is characterized by a half-power semi-angle and center luminous intensity (measured in cd). The receiver is a simple non-imaging photodetector with an optical filter, optical concentrator and a single photodiode element. Its characteristics are as follows: the FOV (field of view) angle, the gain (being a product of filter and concentrator gains), the photodetector area and the conversion efficiency (measured in A/W).

The metric that we use to measure the impact of the rogue transmitter is the Bit Error Rate (BER), which depends on the SNR (Signal to Noise Ratio) and modulation scheme, for M-PAM modulation:

$$BER_{PAM}(SNR) \cong \frac{M-1}{M \log M} Q\left(\sqrt{\frac{SNR}{2(M-1)}}\right) \quad (1)$$

$$BER_s = BER_{PAM}(SNR_s) \quad (2)$$

The approximation in (1) results from the assumption that only one bit errors occur in Gray coding [26].

We calculate SNR as follows:

$$SNR_s = \frac{\overline{s_{data}^2}}{\left(N + \overline{s_{rogue}^2}\right)} \quad (3)$$

where $s_{data}$ is the data signal, $s_{rogue}$ is the signal transmitted by "rogue" (interfering) luminaires, and $N$ is noise.

We want to test the influence of the rogue data source on the legitimate transmitter, as well as the opposite: how the legitimate transmitter influences the signal quality of the rogue one. Hence, we also want to calculate the BER and SNR assuming that the rogue signal is data and that the legitimate signal is treated as noise:

$$BER_r = BER_{PAM}(SNR_r) \quad (4)$$

where

$$SNR_r = \frac{\overline{s_{rogue}^2}}{\left(N + \overline{s_{data}^2}\right)} \quad (5)$$

The problem of noise in VLC environments has been studied in detail [27]. In general, the following noise sources should be considered: background and transmitting LED shot noise, thermal noise in the detector and the influence of ISI (Inter Symbol Interference):

$$N = \sigma_{shot}^2 + \sigma_{thermal}^2 + \sigma_{ISI}^2 \quad (6)$$

where $N$ is the total noise variance and $\sigma_{shot}$, $\sigma_{thermal}$, $\sigma_{ISI}$ is the standard variance of shot, thermal and ISI respectively. Proper estimation of noise in VLC environments is crucial in studying maximum attainable transfer rates under various conditions and modulation schemes. The input-referred noise variance depends on the signal data rate; for data rates of interest in our scenario ($10^5 – 10^7$ bits/s), the dominating noise factors are ISI and background shot noise. Hence, we will consider only these sources:

$$\sigma_{shot}^2 = 2qRPB + 2qI_{bg}I_2B \quad (7)$$

where $q$ is the electronic charge, $R$ is the responsivity of the photodiode, $B$ is the equivalent noise bandwidth, $P$ is the received power, $I_{bg}$ is the background current; and where for a p-i-n/FET receiver, we assume $I_2$ = 0.56. In the multi-luminaire study that we conduct in this paper, the dominant noise factor is the interfering signal from neighboring luminaires and not physical noise itself.

Now we will present an analytical model of the optical wireless channel, which will let us derive SNR and BER measures for different physical scenarios.

A single LED is a Lambertian emitter – its radiation intensity is a cosine function of the viewing angle and is given by:

$$R(\theta) = P_t \frac{(m+1)}{2\pi} \cos^m(\theta) \qquad (8)$$

where $\theta$ is the irradiance angle, $P_t$ is the transmitted power and $m$ is the order of Lambertian emission which is given by the irradiance semi-angle $\theta_{1/2}$ (the half-power angle):

$$m = -\frac{\ln 2}{\ln(\cos(\theta_{1/2}))} \qquad (9)$$

Light propagates from the LED to the receiver via a channel which is modeled by the direct channel transfer function $h_d$:

$$h_d = \begin{cases} \frac{(m+1)A\cos^m(\theta)}{2\pi d^2}\cos(\psi) R(\varphi) & 0 \leq \theta \leq \theta_{FOV} \\ 0 & \theta > \theta_{FOV} \end{cases} \qquad (10)$$

where $\theta$ is the irradiance angle, $\varphi$ is the angle of incidence, $A$ is the receiver's area, $R(\psi)$ is the receiver's gain, $d$ is the distance from the LED to the receiver and $\theta_{FOV}$ is the receiver's field-of-view semi-angle. The geometric model of this simple LOS (Line of Sight) case is shown in Fig. 1.

To make the model more realistic, we will also consider the effect of first reflections from the walls; the channel transfer function for the reflection path – $h_r$ is:

$$h_r = \begin{cases} \frac{(m+1)A\cos^m(\theta)}{2\pi^2 d_1^2 d_2^2} \rho\, dA_w \cos(\alpha)\cos(\beta)\cos(\psi) R(\varphi) & 0 \leq \theta \leq \theta_{FOV} \\ 0 & \theta > \theta_{FOV} \end{cases} \qquad (11)$$

where $\theta$ is the irradiance angle, $\psi$ is the angle of incidence, $A$ is the receiver's area, $R(\varphi)$ is the receiver's gain, $d_1$ is the distance from the LED to a reflective point, $d_2$ is the distance from a reflective point to the receiver, $\rho$ is the reflectance factor, $dA_w$ is a reflective area surface element, $\alpha$ is the angle of irradiance to a reflective point and $\beta$ is the angle of irradiance to the receiver. The geometric model of this case is shown on Fig. 1.

For a single source, the output signal of the LED transmitter is given by the following general formula:

$$p_o(t) = P_t[1 + \mu\, x(t)] \qquad (12)$$

where: $P_t$ is the power transmitted from a single LED, $\mu$ is the modulation index, and $x(t)$ is the modulating signal. Assuming that the receiver is DC blocked, we get the following general formula for the received signal:

$$s(t) = h\, P_t\, \mu\, x(t) \qquad (13)$$

Considering the "legitimate" and "rogue" sets of transmitters, we obtain the following:

$$s_{data}(t) = \sum_{data\_LEDs} \left\{ P_{LED}\, \mu\, x(t)\, h_d + \int_{walls} P_{LED}\, \mu\, x(t) dh_{ref} \right\} \qquad (14)$$

$$s_{rogue}(t) = \sum_{rogue\_LEDs} \left\{ P_{LED}\, \mu\, x(t)\, h_d + \int_{walls} P_{LED}\, \mu\, x(t) dh_{ref} \right\} \qquad (15)$$

We use (14) and (15) in a numeric model to calculate the BER as given in (2) and (4) for our scene.

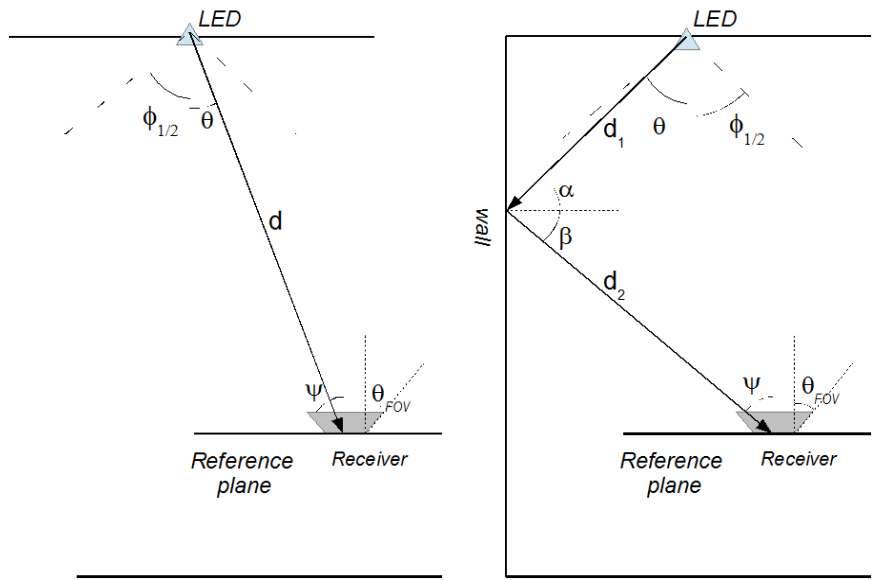

**Fig.** 1 The geometric model of the LOS (left) and NLOS (right) illumination.

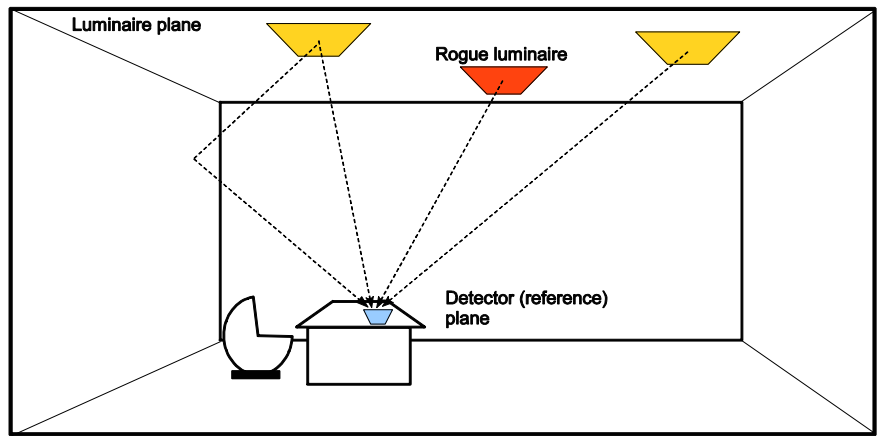

**Fig.** 2 General simulated scene view – both "legal" and "rogue" luminaires are arranged on the luminaire plane.

## 4. The simulated scene

### *4.1. The simulated environment*

For our numerical simulations, we will use a standard "office environment" arrangement, akin to the ones typically used in similar work [1, 21, 22, 23, 24]. However, we have decided to adopt a larger room size than was considered in the above mentioned work, as in our opinion it better fits real-world environments. We have also considered realistic luminaire arrangements used with typical off-the shelf lighting products. We have designed a single room scene illustrated in Fig. 2. The room's dimensions are: 7m x 7 m x 2.8 m. The detector's photodiode at the reference plane is set at the "desk level" – 0.85 m. The room and VLC system's parameters are given in tables 1, 2 and 3. We have tested several different scenarios of luminaire placement with different arrangements of legitimate and rogue transmitters. To design and verify luminaire arrangement, we used a two-step approach: in the first step, a specialized CAD tool [28] was used. This allowed us to set-up realistic scenarios based on standard luminaire components. The scene was initially designed to provide ergonomic lighting conditions which meet current standards[2] [29, 30]. A sample scene as modeled in the CAD tool is shown in Fig. 3. The light power levels, the SNR and BER, for different luminaire placement scenarios were calculated numerically according to formulas (2), (4), (14) and (15) - as described in section 5.

We have simulated three luminaire placement scenarios:

1. **G1** – a 3x4 recessed luminaire grid – Fig. 4,
2. **G2** – a 2x4 recessed luminaire grid with an additional 1x3 array of "downlight" lower power elements – Fig. 5,
3. **GC** – a 2 x 3 luminaire grid surrounded by a circular arrangement of lower power "downlight" elements – Fig. 6.

The first scenario is the most common, standard solution for office spaces. The second scenario is a typical light placement for office spaces (meeting rooms, open spaces, etc.), with downlights providing additional illumination in recessed areas which are further from natural sources of light, such as corridors, etc. The third scenario provides a more uniform light distribution in the room perimeter, which is a favorable feature for VLC communication. In scenario G1, one type of luminaire is used - a square 60 x 60 cm panel module with a 70 deg. radiation semi-angle. In scenarios G2 and GC, two types of luminaires are used: a square 60 x 60 cm panel module with a 70 deg. radiation semi-angle and lower-power luminaires with a narrower radiation semi-angle of 30 deg. In all cases, we have introduced rogue transmitters.

### *4.2. Rogue transmitter choice and placement*

We have tested various scenarios of rogue transmitter placement, taking into account the practical possibility of an attacker modifying the VLC infrastructure. We have only considered scenarios in which a whole set of LEDs constituting a luminaire is taken over by an attacker. This is a realistic assumption taking into account the fact that all LEDs in the luminaire are driven by a single amplifier and modulator, and it is not technically feasible to hack "a part" of the luminaire. Scenario G1 corresponds to the case where one of the luminaires was physically modified or replaced (Fig. 4) – one out of 12 luminaires is a rogue one. Two different rogue transmitter placement options were tested here – central and peripheral. In scenario G2 – Fig. 5, we considered the possibility of an attack where one or all 3 of the downlight luminaires are taken over. In scenario GC – Fig. 6, we tested the possibility of taking over all 10 (the full "circle") or 5 (half of the "circle") of the downlight luminaires.

**Table 1** – Physical parameters of the simulated scenarios

| Photodetector parameters | |
|---|---|
| FOV (field of view) | 60 $^o$ |
| Detector area | 1 cm$^2$ |
| Detector gain | 4.5 |
| **Room parameters** | |

---

[2] Although some controversy remains as to the proper light-level standards for office workers, we have assumed an optimal level of 500 lx and not less than 300 lx for the desk surface.

| Dimensions | 7m x 7 m x 2.8 m |
|---|---|
| Reference plane height | 0.85 m |
| Wall reflectivity coefficient | 0.8 |

**Table 2** – Physical parameters of the simulated scenarios – the luminaire arrangement

| Scene type: | G1 | G2 | GC |
|---|---|---|---|
| Scene description: | 3 x 4 grid | 2 x 4 grid + 1 x 3 | 2 x 3 grid + circle |
| horizontal luminaire spacing | 180 cm | 150 - 180 cm | 240 cm |
| vertical luminaire spacing | 150 cm | 180 – 240 cm | 180 cm |
| # luminaires type "g" | 12 | 8 | 6 |
| # luminaires type "d" | - | 3 | 10 |
| Rogue luminaire arrangement | 1 type g | 1, 3 type d | 1,5,10 type d |

**Table 3** – Physical parameters of the LEDs and luminaires

| Luminaire type: | type "g" | type "d" |
|---|---|---|
| Dimensions | 60 x 60 cm | 20 x 20 cm |
| LED irradiance semi-angle at half power | 70 deg | 30 deg (45, 50 deg) |
| Luminous flux | 2000 lm | 1600 lm |
| # of LEDs | 36 | 9 |
| LED placement | 6x6 square grid, 10 cm spacing | 3x3 square grid, 5 cm spacing |

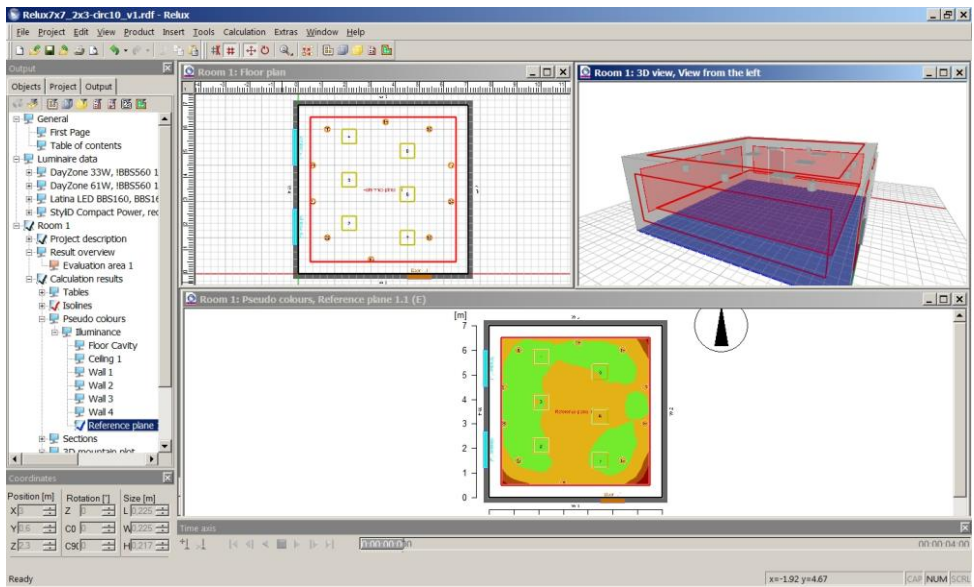

**Fig.** 3 – Screenshot from the "Relux" CAD tool used in the planning of the luminaire layout.

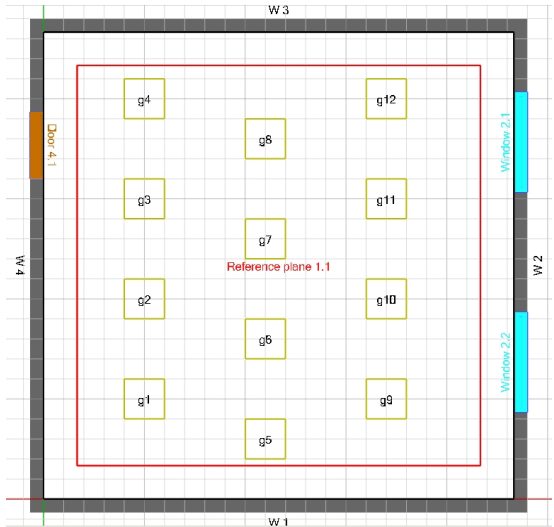

**Fig.** 4 – Scene G1, luminaire arrangement - the rogue luminaires in two sub-scenarios are g7 or g10.

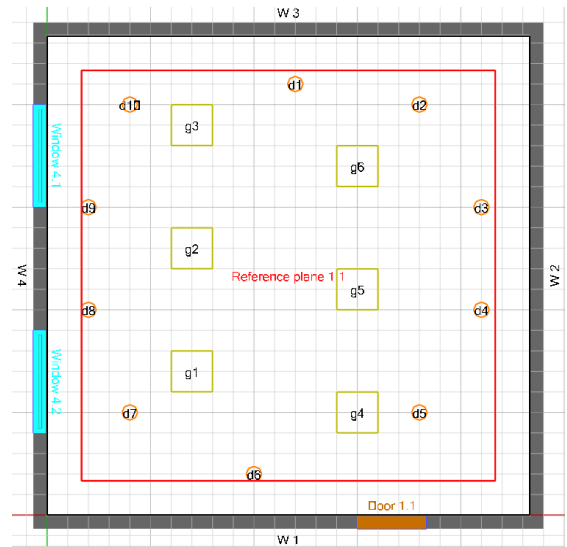

**Fig.** 6 – Scene GC, luminaire arrangement - the rogue luminaires in two sub-scenarios are d1-d10 or d1-d5.

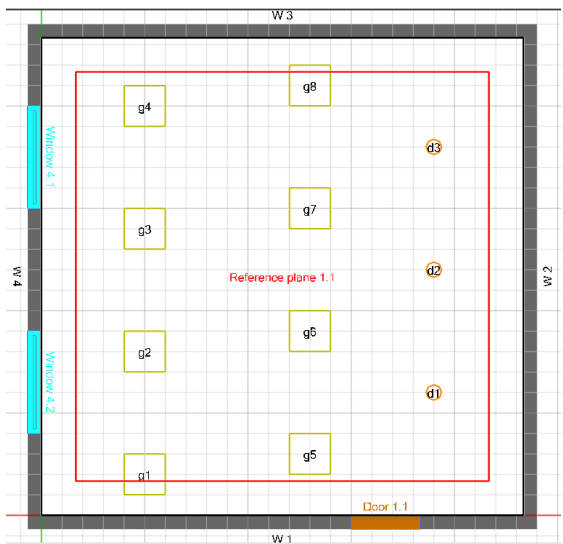

**Fig.** 5 – Scene G2, luminaire arrangement - the rogue luminaires in two sub-scenarios are d2 or d1-d3.

## 5. Simulation results

In all the scenarios, we show the logarithmic plots of the computed Bit Error Rate achievable for legitimate and rogue data transmission: $BER_s$ (2) and $BER_r$ (4). We assume that a maximum BER level of $10^{-3}$ is required for effective transmission.

Scene G1 – in this scenario, we test an arrangement with one rogue transmitter placed in two locations: centrally and peripherally. With the rogue placed centrally, legitimate transmission is jammed in 11% of the total room area – Fig. 7; with peripheral placement, legitimate transmission is jammed in 18% of the area – Fig. 8. With the $BER_r$ not lower than $10^{-1}$, effective rogue transmission is not possible in this scenario – Fig. 9, 10.

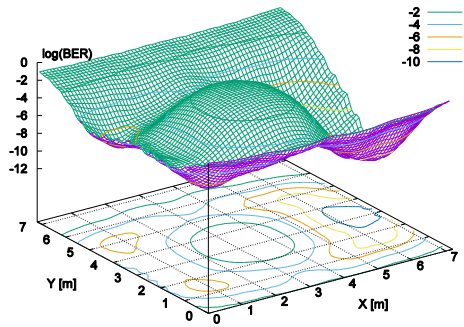

**Fig.** 7 – Simulation results: BER for scenario G1 – central rogue placement.

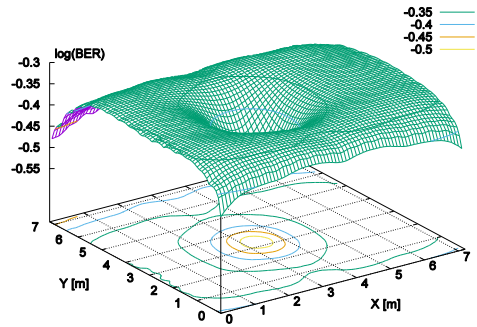

**Fig.** 9 – Simulation results: BER for rogue data, scenario G1 – central rogue placement.

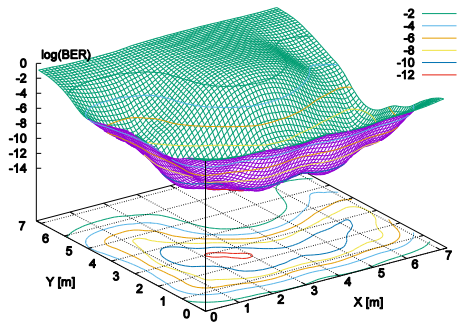

**Fig.** 8 – Simulation results: BER for scenario G1 – peripheral rogue placement.

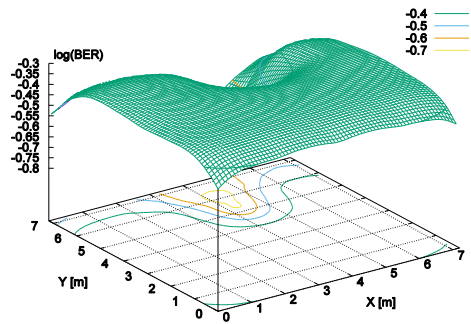

**Fig.** 10 – Simulation results: BER for rogue data, scenario G1 – peripheral rogue placement.

Scene G2 – in this scenario, we test two sub-arrangements: with one and with three rogue transmitters. In the case of one rogue transmitter placed centrally in the downlight row – legitimate transmission is jammed in 10% of the total room area – Fig. 11. Rogue transmission is possible in a very limited area directly under the transmitter (less than 1% of the total area) – Fig. 12. With three rogue downlight luminaires legitimate transmission is jammed in an area of 42% – Fig. 13, while rogue transmission is possible in an area of 10% – Fig. 14.

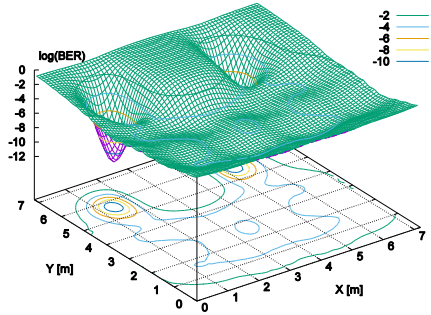

**Fig.** 11 – Simulation results: BER for scenario G2 – one rogue luminaire.

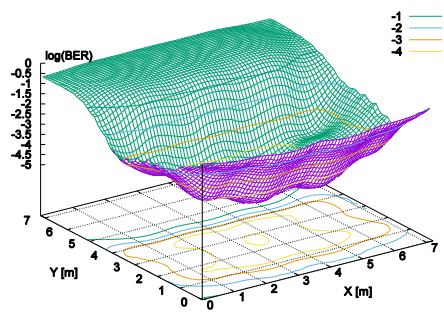

**Fig.** 13 – Simulation results: BER for scenario G2 – three rogue luminaries.

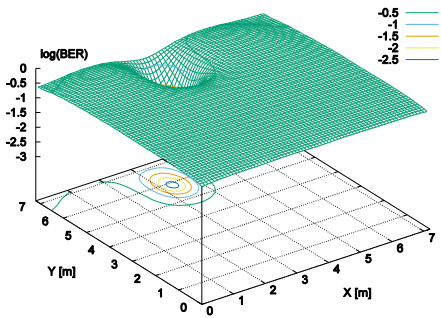

**Fig.** 12 – Simulation results: BER for rogue data, scenario G2 – one rogue luminaire.

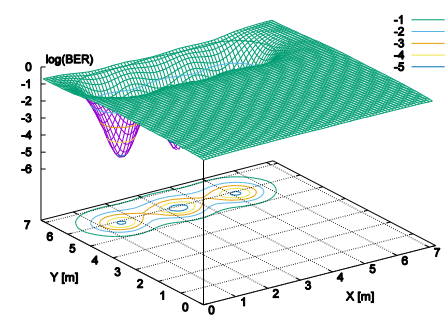

**Fig.** 14 – Simulation results: BER for rogue data, scenario G2 – three rogue luminaries.

In this scenario we have also tested the impact of an irradiance semi-angle on the BER, increasing $\varphi_{1/2}$ from the default value $30^o$ deg. to $45^o$, for all type "d" (rogue) luminaires (transmitted power remains constant). As the semi-angle is increased, the $BER_r$ increases to values in the range of $10^{-1}$, making rogue transmission infeasible – Fig. 15 and Fig. 16. With an increase in the rogue transmitter semi-angle, the area of jammed legitimate transmission (BERs above $10^{-3}$) increases from 10% to 25% (comparing with results shown on Fig. 11).

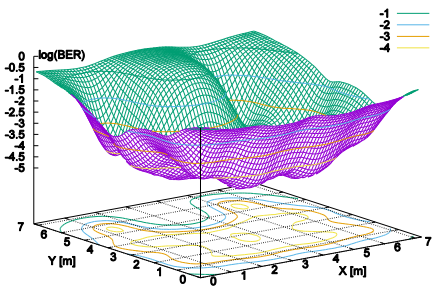

**Fig. 15** - Simulation results: BER for scenario G2, semi-angle for "d" luminaires increased to 45 deg.

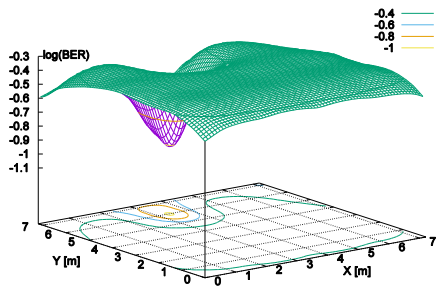

**Fig. 16** - Simulation results: BER for rogue data for scenario G2, semi-angle for "d" luminaires increased to 45 deg.

Scene GC – in this scenario, we also test two sub-scenarios: with 5 and 10 rogue transmitters. In the first case, half of downlight luminaire "circle" is rogue; in the second sub-scenario, whole downlight "circle" consists of rogue luminaires. With 5 rogue luminaires, legitimate transmission is possible in an area of 50%, while rogue transmission again is attainable only in small recessed areas (less than 2.5% of the total area) – Fig 17,18. With 10 rogue luminaires on the perimeter, jamming is very effective -legitimate transmission is possible only in an area of 25% – Fig. 19, while rogue transmission is feasible only in the recessed room areas (less than 5% of the total area) – Fig. 20.

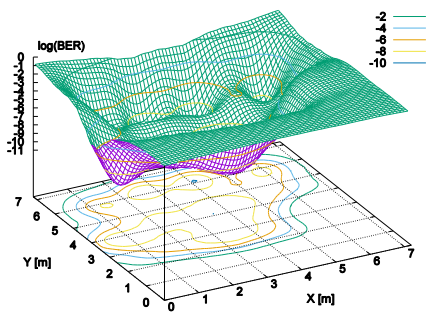

**Fig.** 17 – Simulation results: BER for scenario GC – 5 out of 10 luminaires in the circular arrangements are rogue.

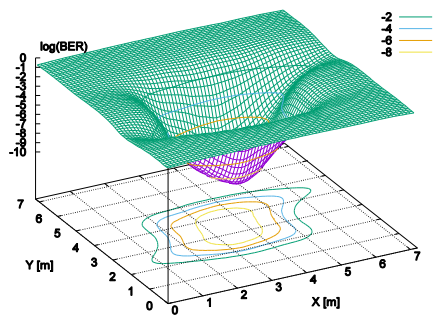

**Fig.** 19 – Simulation results: BER for scenario GC – all luminaires in the circular arrangements are rogue.

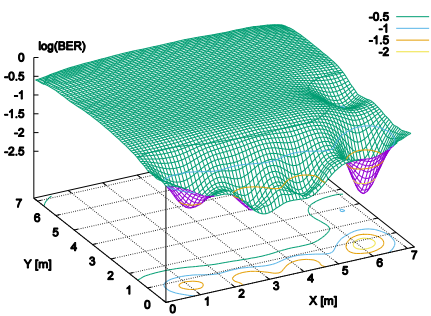

**Fig.** 18 – Simulation results: BER for rogue data, scenario GC – 5 out of 10 luminaires in the circular arrangements are rogue

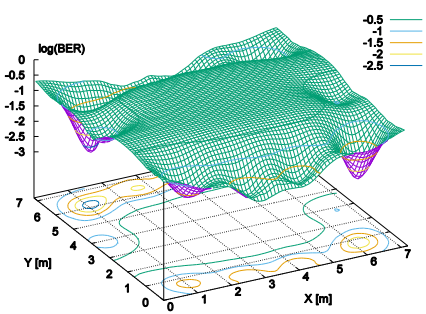

**Fig.** 20 – Simulation results: BER for rogue data, scenario GC – all luminaires in the circular arrangements are rogue

## 6. Summary

VLC networks are currently considered as a means of augmenting traditional radio-based WLANs. However, the issues of VLC security are only beginning to be tackled. As with radio-based networks, the shared nature of optical medium allows adversaries to generate jamming and hijacking attacks. In this work, we have presented a numerical study of a VLC system with rogue transmitters. To the best of our knowledge, it is the first time the problem of interfering legitimate and illegal VLC transmission has been considered. Our simulation "scenes" have been designed and verified with a CAD planning tool to ensure relevance to real-world environments. We have numerically calculated the Bit Error Rates of interfering legitimate and rogue transmission in various office room scene arrangements. We have concluded that it is feasible, with limited access to the transmitter infrastructure, to jam transmission in a large area of up to 40% of the whole scene. It is also possible to introduce illegal transmission on limited areas of up to 10% of the whole scene. In our opinion, this study brings forward the issue with the security of VLC based wireless LANs with respect to rogue transmission. As WYSIWYS – "What You See Is What You Send" remains mostly a slogan, appropriate security mechanisms must be implemented in the data-link layer (and/or higher network layers) to cope with the possibility of rogue transmissions. Detecting the presence of a jamming attack in VLC networks also remains an open issue. In [31], various "consistency checking" methods for detecting a wireless radio jammer have been proposed; the adoption of these techniques for VLC systems and the development of new rogue transmitter detection methods dedicated to VLC would also be interesting areas for future research.

**Acknowledgement**

This work was supported by the Statutory Grant of the Polish Ministry of Science and Higher Education, given to the Institute of Computer Science, Warsaw University of Technology.

**Bibliography**

[1] M. Nakagawa, "Visible Light Communications," In Proc. Conference on Lasers and Electro-Optics/Quantum Electronics and Laser Science Conference and Photonic Applications Systems Technologies, Baltimore, 2007, DOI: 10.1109/CCNC.2012.6181092

[2] H. Elgala, R. Mesleh and H. Haas, "Indoor Optical Wireless Communication: Potential and State-of-the-Art," IEEE Communications Magazine, Volume: 49, Issue: 9, 2011, pp. 56-62.

[3] A. Tsiatmas, C.P. A. Baggen, F.M. Willems, J.P. Linnartz and J.W. Bergmans, "An illumination perspective on visible light communications," In Communications Magazine, IEEE, 52.7, 2014, pp. 64-71.

[4] Samsung Electronics, ETRI, VLCC, University of Oxford, "Visible Light Communication: Tutorial," 2008, http://www.ieee802.org/802_tutorials/2008-03/15-08-0114-02-0000-VLC_Tutorial_MCO_Samsung-VLCC-Oxford_2008-03-17.pdf

[5] M. B. Rahaim, A.M. Vegni and T. D. Little, "A hybrid radio frequency and broadcast visible light communication system," in Proc. IEEE Global Communications Conference (GLOBECOM) Workshops, 2011, pp. 792–796.

[6] L.B. Chen, et al. "Development of a dual-mode visible light communications wireless digital conference system," In Consumer Electronics (ISCE 2014), The 18th IEEE International Symposium on, 2014, pp. 1-2.

[7] J. P. Javaudin, M. Bellec, D. Varoutas and V. Suraci, "OMEGA ICT Project: Towards Convergent Gigabit Home Networks," in Proc. International Symposium on Personal, Indoor and Mobile Radio Communications (PIMRC), Cannes, France, 2008

[8] K.D. Langer, et al., "Optical Wireless Communications for Broadband Access in Home Area Networks," In Proc. International Conference on Transparent Optical Networks, ICTON,

2008, pp. 149 - 154, DOI: 10.1109/ICTON.2008.4598756

[9] D.C. O'Brien, et al, "Home access networks using optical wireless transmission," In Proc. Personal, Indoor and Mobile Radio Communications, 2008, IEEE 19th International Symposium on, pp. 1-5, 2008

[10] D.C. O'Brien, et al, "Gigabit Optical Wireless for a Home Access Network," in Proc. IEEE 20th International Symposium on Personal, Indoor and Mobile Radio Communications, 2009, pp. 1-5.

[11] M. Yoshino, S. Haruyama and M. Nakagawa, "High-accuracy positioning system using visible LED lights and image sensor," Radio and Wireless Symposium, IEEE, vol., no., 2008, pp.439-442, 22-24.

[12] Z. X. Ren, H. M. Zhang, L. Wei and Y. Guan, "A High Precision Indoor Positioning System Based on VLC and Smart Handheld," in Applied Mechanics and Materials, Vol. 571, 2014, pp. 183-186.

[13] GBI Research, "Visible Light Communication (VLC) - A Potential Solution to the Global Wireless Spectrum Shortage," Tech. Rep. GBI Research, 2011,

[14] Home Gigabit Access (OMEGA) Project. [Online]. Available: http://www.ict-omega.eu/

[15] Q. Wang, D. Giustiniano and D. Puccinelli, D., "OpenVLC: Software-defined visible light embedded networks," In Proceedings of the 1st ACM MobiCom workshop on Visible light communication systems, September 2014, pp. 15-20

[16] J.P. Conti, "What you see is what you send," Engineering & Technology, 2008, pp. 66-67.

[17] G. Blinowski, "Practical aspects of physical and MAC layer security in visible light communication systems,". *International Journal of Electronics and Telecommunications*, 2016, *62*(1), pp. 7-13.

[18] IEEE, "IEEE standard for local and metropolitan area networks–part 15.7: Short-range wireless optical communication using visible light", IEEE Std 802.15.7-2011, https://standards.ieee.org/findstds/standard/802.15.7-2011.html

[19] J. Classen, J. Chen, J., D. Steinmetzer, M. Hollick, and E. Knightly, "The Spy Next Door: Eavesdropping on High Throughput Visible Light Communications'" In Proceedings of the 2nd ACM MobiCom Workshop on Visible Light Communication Systems, ser. VLCS (Vol. 15), 2015

[20] A. D. Wyner, "The wire-tap channel," The Bell System Technical Journal, vol. 54, pp. 1355–1387, 1975.

[21] H. Le Minh, A. T. Pham, Z. Ghassemlooy and A. Burton, "Secured Communications-Zone Multiple Input Multiple Output Visible Light Communications," In Proc. Globecom Workshop - Optical Wireless Communications, 2014

[22] A. Mostafa and L. Lampe, "Physical-Layer Security for Indoor Visible Light Communications," In Proc. IEEE ICC 2014 - Optical Networks and Systems

[23] C.-W. Chow, "Secure communication zone for white-light LED visible light communication,", in Optics Communications 344, pp. 81–85; 2015

[24] K. Cui, J. Quan and Z. Xu, "Performance of indoor optical femtocell by visible light communication," Optics Communications, 2013, pp. 59-66.

[25] M. Wilhelm, I. Martinovic, J. B. Schmitt and V. Lenders, "Reactive jamming in wireless networks: how realistic is the threat?" In Proc. of the fourth ACM conference on Wireless network security, pp. 47-52, ACM, 2011

[26] F. Xiong, Digital Modulation Techniques, Second Edition, Boston, MA: Artech House, 2006, ISBN: 978-1-58053-864-0

[27] A. Agarwal and S. Garima, "SNR Analysis for Visible Light Communication Systems," International Journal of Engineering Research and Technology. Vol. 3. No. 10 (October-2014). ESRSA Publications, 2014.

[28] Relux Suite, Relux Informatik AG, http://www.relux.biz

[29] EU Standard EN 12464-1:2011, Light and lighting. Lighting of work places. Indoor work places, 2011

[30] M. Wright, "Philips Lighting questions proper light-level standards for office workers", LEDs Magazine and Illumination in Focus, 2015, http://www.ledsmagazine.com/articles/iif/2015/03/philips-lighting-questions-proper-light-level-standards-for-office-workers.html

[31] Xu, Wenyuan, et al. "The feasibility of launching and detecting jamming attacks in wireless networks." *Proceedings of the 6th ACM international symposium on Mobile ad hoc networking and computing*. ACM, 2005.